\documentclass[review,12pt]{elsarticle}
\usepackage{amsmath,amsthm,amsfonts,amssymb,mathtools,lineno}
\usepackage[hidelinks]{hyperref}
\hypersetup{colorlinks, linkcolor=blue, citecolor=blue}
\newtheorem*{remark}{Remark}
\newcommand{\caputo}[1]{{}^{C}\!D_{t}^{\beta }#1\left( t \right)}
\journal{arXiv.org}

\begin{document}

\begin{frontmatter}

\title{Modified Fractional Logistic Equation}

\author{Mirko D'Ovidio }
\author{Paola Loreti}
\author{Sima Sarv Ahrabi \corref{cor1}}
\ead{sima.sarvahrabi@sbai.uniroma1.it}
\cortext[cor1]{Corresponding author}
\address{Dipartimento di Scienze di Base e Applicate per l'Ingegneria, Sapienza Universit\`{a} di Roma, Via Antonio Scarpa n. 16, 00161 Rome, Italy}

\begin{abstract}
In the article [B. J. West, Exact solution to fractional logistic equation, Physica A: Statistical Mechanics and its Applications 429 (2015) 103--108], the author has obtained a function as the solution to fractional logistic equation (FLE). As demonstrated later in [I. Area, J. Losada, J. J. Nieto, A note on the fractional logistic equation, Physica A: Statistical Mechanics and its Applications 444 (2016) 182--187], this function (West function) is not the solution to FLE, but nevertheless as shown by West, it is in good agreement with the numerical solution to FLE. The West function indicates a compelling feature, in which the exponentials are substituted by Mittag--Leffler functions. In this paper, a modified fractional logistic equation (MFLE) is introduced, to which the West function is a solution. The proposed fractional integro--differential equation possesses a nonlinear additive term related to the solution of the logistic equation (LE). The method utilized in this article, may be applied to the analysis of solutions to nonlinear fractional differential equations of mathematical physics.

\end{abstract}

\begin{keyword}
Fractional differential equations \sep logistic equation \sep asymptotic expansions \sep Mittag-Leffler functions \sep subordinators.
\MSC 26A33 \sep 34A08 \sep 35B40.
\end{keyword}
\end{frontmatter}

\section{Logistic equation}\label{sec.logistic}

The logistic equation, which is mentioned on occasion as the \emph{Verhulst} model, is a population growth model introduced and published by Pierre Verhulst \cite{verhulst1838}. The model represents a well-known nonlinear differential equation in the field of biology and social sciences:
\begin{equation}\label{eq.Verhul}
\frac{\mathrm{d}N\left( t \right)}{\mathrm{d}t}=kN\left( t \right)\left( 1-\frac{1}{{{N}_{\max }}}N\left( t \right) \right)\,,\qquad t\ge 0\, ,
\end{equation}
where $k$ is the rate of maximum population growth constrained to be a real positive number,  $N\left( t \right)$ is the population and ${{N}_{\max }}$ is the carrying capacity, i.e.\ the maximum attainable value of population. By dividing both side of Eq. \eqref{eq.Verhul} by ${{N}_{\max }}$ and defining $u=N\left( t \right)/{N}_{\max }$ as the normalization of population to its maximum sustainable value, the differential equation
\begin{equation}\label{eq.OrdinLogis}
\frac{\mathrm{d}u}{\mathrm{d}t}= k u \left( 1-u \right) \,, \qquad t\ge 0 \,,
\end{equation}
is obtained, for which there is an exact closed form solution
\begin{equation}\label{eq.LogisSol}
u\left( t \right) = \frac{{{u}_{0}}}{{{u}_{0}}+\left( 1-{{u}_{0}} \right){{e}^{-kt}}} \,,\qquad t\ge 0\, ,
\end{equation}
where ${{u}_{0}}$ is the initial state at the time $t=0$. The sigmoidal behavior of the solution to the logistic equation has been also used to model the tumor growth \cite{forys2003} and so forth. Since the logistic growth is one of the most versatile models in natural sciences, the fractional logistic equation would be a relevant problem to be dealt with.

The Laplace transform method cannot directly lead up to a solution of such a nonlinear fractional differential equation. In \cite{das2010,yang2015}, the authors represented some creative techniques to approximate the solution to FLE. The authors of the article \cite{kumar17analys} have analysed the FLE in the sense of a recently defined fractional derivative, which is mentioned as Caputo--Fabrizio fractional derivative \cite{caputo15new}, and represented the solution by utilizing numerical methods. In \cite{ortig17new}, the authors have studied the FLE with the Gr\"{u}nwald--Letnikov fractional derivative and assumed the solution to be in the form of a fractional Taylor series, where the coefficients in the series are evaluated by a recursive relation. The Carleman embedding technique has been employed by Bruce J. West (see \cite{west2015}) to construct a solution to fractional logistic equation
\begin{equation}\label{eq.FracLogis}
\caputo{w}={{k}^{\beta }}w\left( 1-w \right) \,, \qquad \beta \in \left(0,1\right] \,,
\end{equation}
with the initial condition $w \left(0\right) = u_0$, where ${}^{C}\!D_{t}^{\beta }$ denotes the Caputo fractional differential operator with the fractional order, $\beta$, restricted to $0<\beta \le 1$. The proposed solution, which has been obtained by West \cite{west2015} and is mentioned as West function (WF), is
\begin{equation}\label{eq.SoluWest}
w\left( t \right)=\sum\limits_{n=0}^{\infty }{{{\left( \frac{{{u}_{0}}-1}{{{u}_{0}}} \right)}^{n}}{{E}_{\beta }}\left( -n{k^\beta}{{t}^{\beta }} \right)} \,, \qquad \beta \in \left(0,1\right] \,,
\end{equation}
where $E_\beta$ denotes the so--called one parameter Mittag--Leffler function, but nonetheless in \cite{area2016}, the authors have illustrated that the WF is not the solution to fractional differential equation \eqref{eq.FracLogis} except the case, where the fractional order, $\beta$, is equal to one. However, as demonstrated in \cite{west2015}, the WF has been shown to be in good agreement with the numerical solution of the FLE.

The discussion on the FLE is motivated by the relevance of the model to a wide range of applications and by the mathematical difficulties involved in the analysis of nonlinear fractional equations emerging in mathematical biology. The aim of this article is to investigate what equation may be satisfied by the WF (for the case $k=1$), i.e.\ the goal is to seek for an equation which could be satisfied by
\begin{equation}\label{eq.Solution}
w\left( t \right)=\sum\limits_{n=0}^{\infty }{{{\left( \frac{{{u}_{0}}-1}{{{u}_{0}}} \right)}^{n}}{{E}_{\beta }}\left( -n{{t}^{\beta }} \right)}\, ,
\end{equation}
In this regard, the fractional integro--differential equation
\begin{eqnarray}\label{eq.Main}
  \caputo{w} &=& w\left( t \right)\left( 1-w\left( t \right) \right)+{{u}_{0}}\frac{{{t}^{-\beta }}}{\Gamma \left( 1-\beta  \right)} \nonumber \\
  & &+ \int_{0}^{\infty }\!\!\!\!{\int_{0}^{\infty }\!\!\!{\left( u\left( s \right)u\left( z \right)-{{u}^{2}}\left( s \right) \right){{l}_{\beta }}\left( s,t \right)}}{{l}_{\beta }}\left( z,t \right)\mathrm{d}s \, \mathrm{d}z \, ,
\end{eqnarray}
with the initial condition $w \left( 0 \right) = u_0$ is represented and proved to be satisfied by the function described in \eqref{eq.Solution}. In Eq. \eqref{eq.Main}, which is called as modified fractional logistic equation (MFLE), the function $u$ is the solution to the logistic equation \eqref{eq.OrdinLogis} for the case $k = 1$. Thus, Eq. \eqref{eq.Main} has an additive term related to the solution of the classical logistic equation. The function $l_\beta \left(s,t\right)$ is the unique solution to the equation
\begin{equation}\label{eq.DelLbeta}
  {}^{C}\!D_{t}^{\beta }{{l}_{\beta }}\left( s,t \right)=-\frac{\partial }{\partial s}{{l}_{\beta }}\left( s,t \right) \, ,
\end{equation}
with the initial condition
\begin{equation}\label{eq.inilbeta}
  {{l}_{\beta }}\left( s,0 \right)=\delta \left( s \right) \, ,
\end{equation}
where $\delta \left( s \right)$ stands for the Dirac's delta function, and the boundary condition
\begin{equation}\label{eq.boundarylbeta}
  {{l}_{\beta }}\left( 0,t \right)=\frac{{{t}^{-\beta }}}{\Gamma \left( 1-\beta  \right)} \,,
\end{equation}
and furthermore
\begin{equation}\label{eq.ProbInvInf}
\int_{0}^{\infty }\!\!\!{{{l}_{\beta }}\left( s,t \right)\mathrm{d}s}=1\,.
\end{equation}
The Laplace transform of the function $l_\beta \left(s,t\right)$ is
\begin{equation}\label{eq.MittagDistribution}
\int_{0}^{\infty }\!\!\!{{{e}^{-\lambda s}}{{l}_{\beta }}\left( s,t \right)\mathrm{d}s}={{E}_{\beta }}\left( -\lambda {{t}^{\beta }} \right) \, , \qquad \lambda>0 \,.
\end{equation}
Further details about ${{l}_{\beta }}\left( s,t \right)$ can be observed in, for instance, \cite{bingham1971,bertoin1999subordinators}.

Some necessary preliminaries about asymptotic behaviour of Mittag--Leffler function, ${{E}_{\beta }}\left( z \right)$, will be briefly discussed in section \ref{sec.mittag}. Section \ref{sec.ModifiedLogistic} is entirely devoted to the solution of the fractional integro--differential equation \eqref{eq.Main}, and fractional order estimation of which will be discussed in section \ref{sec.estimation}.

\section{Mittag--Leffler function}\label{sec.mittag}

The so-called one parameter Mittag-Leffler function ${{E}_{\beta }}\left( z \right)$ is defined as a power series, denoted by
\begin{equation}\label{eq.Mittag}
{{E}_{\beta }}\left( z \right)=\sum\limits_{k=0}^{\infty }{\frac{{{z}^{k}}}{\Gamma \left( \beta k+1 \right)}}\, ,\qquad \beta >0\, \qquad z \in \mathbb{C} \,.
\end{equation}
which was first introduced by G. M. Mittag-Leffler and could be considered as the generalization of the exponential function due to the replacement of $\Gamma \left( k+1 \right)$ by $\Gamma \left( \beta k+1 \right)$ in the exponential series formula (for instance, see \cite{goren16mittag,podlubny1998}). It could be obviously perceived that ${{E}_{\beta }}\left( 0 \right)=1$. In this note, the main focus of attention will be the function
\begin{equation}\label{eq.MittagMinus}
{{E}_{\beta }}\left( -\lambda {{z}^{\beta }} \right)=\sum\limits_{k=0}^{\infty }{{{\left( -1 \right)}^{k}}{{\lambda }^{k}}\frac{{{z}^{k\beta }}}{\Gamma \left( k\beta +1 \right)}} \, ,
\end{equation}
which provides the Laplace transform of $l_\beta$ (see Eq. \eqref{eq.MittagDistribution}). It is appropriately pointed out that the asymptotic behaviour of the Mittag-Leffler function ${{E}_{\beta }}\left( \lambda {{z}^{\beta }} \right)$, for $0<\beta <2$ and $z\in {{\mathbb{R}}^{+}}$, could be stated as follows \cite{podlubny1998,gerhold2012asymptotics,wang2017note,gorenflo1997}
\begin{eqnarray}\label{eq.AsympPos}
  {{E}_{\beta }}\left( \lambda {{z}^{\beta }} \right) &=& \frac{1}{\beta }\exp \left( \frac{z}{{{\lambda }^{\beta }}} \right)-\sum\limits_{k=1}^{n}{\frac{{{z}^{-k\beta }}}{{{\lambda }^{k}}\Gamma \left( 1-k\beta  \right)}}\nonumber\\
  & &+O\left( {{\left| \lambda {{z}^{\beta }} \right|}^{-1-n}} \right)\, ,\quad n\in \mathbb{N}\,, \lambda>0\,, z\to+\infty \,,
\end{eqnarray}
and
\begin{eqnarray}\label{eq.AsympNeg}
  {{E}_{\beta }}\left( \lambda {{z}^{\beta }} \right) &=& -\sum\limits_{k=1}^{n}{\frac{{{z}^{-k\beta }}}{{{\lambda }^{k}}\Gamma \left( 1-k\beta  \right)}}\nonumber\\
  & &+O\left( {{\left| \lambda {{z}^{\beta }} \right|}^{-1-n}} \right)\, ,\quad n\in \mathbb{N} \,, \lambda<0 \,, z\to+\infty \,.
\end{eqnarray}
Furthermore the following inequality is held true for all non-negative real numbers, i.e. $z\in [0,\infty )$ (e.g., see \cite[theorem 1.6]{podlubny1998}):\\
For $0 < \beta < 2$, there exists a constant $C \left( \beta \right)$ such that
\begin{equation}\label{eq.MittagInequal}
0\le \left| {{E}_{\beta }}\left( -{{z}^{\beta }} \right) \right|\le \frac{C\left( \beta \right)}{1+{{z}^{\beta }}}\, , \qquad 0<\beta <2 \, ,
\end{equation}
where $C\left( \beta \right)$ is a real positive constant.

By the Riemann-Liouville fractional order derivative $D_{z}^{\beta }$, the following formula could be obtained (see \cite[formula 2.2.53]{kilbas2006}) for the fractional derivative of Mittag-Leffler function
\begin{equation}\label{eq.MittagDeriv}
D_{z}^{\beta }{{E}_{\beta }}\left( \mu {{z}^{\beta }} \right)=\frac{{{z}^{-\beta }}}{\Gamma \left( 1-\beta  \right)}+\mu {{E}_{\beta }}\left( \mu {{z}^{\beta }} \right) \,,\quad \beta \in{{\mathbb{R}}^{+}} \,,\quad \mu \in \mathbb{C}\,.
\end{equation}
\section{Modified fractional logistic equation}\label{sec.ModifiedLogistic}

\begin{figure}[ht!]
\centering
\includegraphics[scale=.6]{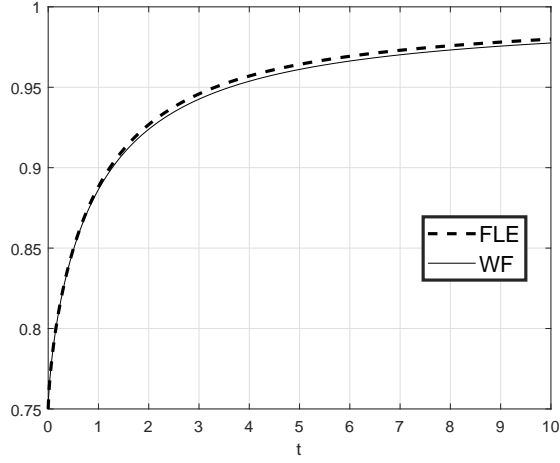}
\caption{Comparison of the West Function (WF) expressed in Eq. \eqref{eq.Solution} and the numerical integration of the FLE (Eq. \eqref{eq.FracLogisticKOne}), for  $\beta = 0.7$ and $u_0=0.75$\,.}
\label{fig1}
\end{figure}

In \cite{west2015}, the author has utilized the Carleman embedding technique to construct an infinite--order system of linear fractional differential equations equivalent to the nonlinear fractional differential equation \eqref{eq.FracLogis} and has obtained a solution in terms of a weighted sum over the Mittag--Leffler functions (see Eq. \eqref{eq.SoluWest}). The authors in \cite{area2016} indicated later that the Carleman embedding technique solves integer--order differential equations, not the fractional ones. Nonetheless, for $\beta=1$, the WF results in the solution to classical logistic equation. As illustrated in Figure \ref{fig1}, it is observed that the WF in \eqref{eq.Solution}, is in good agreement with the numerical integration of fractional logistic equation
\begin{equation}\label{eq.FracLogisticKOne}
\caputo{w}=w\left( 1-w \right) \,, \qquad \beta \in \left(0,1\right] \,.
\end{equation}
Figure \ref{fig1} shows the graph of the numerical solution to the fractional logistic equation \eqref{eq.FracLogisticKOne} and the WF represented in \eqref{eq.Solution} for the fractional order $\beta=0.7$. The MATLAB code fde12.m \cite{garrappa2011predictor}, which implements the predictor--corrector method of Adams-Bashforth-Moulton type described in \cite{diethelm1998fracpece}, is used in order to represent the numerical solution of Eq. \eqref{eq.FracLogisticKOne}. The WF is numerically evaluated by means of the MATLAB code ml.m \cite{garrappa2014mittag}, which is based on the numerical inversion of the Laplace transform of Mittag--Leffler function \cite{garrappa2015numerical}.
In this section, the goal is to demonstrate that the WF, which has been expressed in \eqref{eq.Solution}, is the solution to fractional integro--differential equation \eqref{eq.Main}:\\
Referring to Eqs. \eqref{eq.OrdinLogis} and \eqref{eq.LogisSol} , it can be observed that the solution to classical logistic equation $\dot{u}= u \left( 1-u \right)$ is as follows
\begin{equation}\label{eq.LogisSolKone}
u\left( t \right) = \frac{{{u}_{0}}}{{{u}_{0}}+\left( 1-{{u}_{0}} \right){{e}^{-t}}} \,,\qquad t\ge 0\, ,
\end{equation}
The function $u$ in \eqref{eq.LogisSolKone} can be rewritten as below
\begin{align}\label{eq.LogisSolSum}
  u\left( t \right)&=\frac{{{u}_{0}}}{{{u}_{0}}+\left( 1-{{u}_{0}} \right){{e}^{-t}}}\nonumber\\
  &=\sum\limits_{k=0}^{\infty }{{{\left( \frac{{{u}_{0}}-1}{{{u}_{0}}} \right)}^{k}}{{e}^{-kt}}}\, .
\end{align}
By using Eqs. \eqref{eq.MittagDistribution} and \eqref{eq.LogisSolSum}, the function $w\left(t\right)$, represented in \eqref{eq.Solution}, can be appropriately expressed in terms of ${{l}_{\beta }}\left( s,t \right)$
\begin{align}\label{eq.SolutionDens}
  w\left( t \right)&=\sum\limits_{k=0}^{\infty }{{{\left( \frac{{{u}_{0}}-1}{{{u}_{0}}} \right)}^{k}}{{E}_{\beta }}\left( -k{{t}^{\beta }} \right)}\nonumber\\
  &=\sum\limits_{k=0}^{\infty }{{{\left( \frac{{{u}_{0}}-1}{{{u}_{0}}} \right)}^{k}}\int_{0}^{\infty }\!\!\!{{{e}^{-ks}}{{l}_{\beta }}\left( s,t \right)\mathrm{d}s}}\nonumber\\
  &=\int_{0}^{\infty }\!{\sum\limits_{k=0}^{\infty }{{{\left( \frac{{{u}_{0}}-1}{{{u}_{0}}} \right)}^{k}}{{e}^{-ks}}}{{l}_{\beta }}\left( s,t \right)\mathrm{d}s}\nonumber\\
  &=\int_{0}^{\infty }\!\!\!{u\left( s \right){{l}_{\beta }}\left( s,t \right)\mathrm{d}s}\,.
\end{align}
From equation \eqref{eq.Solution}, it could be obtained that
\begin{align}\label{eq.SolPowerTwo}
  {{w}^{2}}\left( t \right)&=\sum\limits_{k=0}^{\infty }{\sum\limits_{i=0}^{\infty }{{{\left( \frac{{{u}_{0}}-1}{{{u}_{0}}} \right)}^{k+i}}{{E}_{\beta }}\left( -i{{t}^{\beta }} \right){{E}_{\beta }}\left( -k{{t}^{\beta }} \right)}}\nonumber\\
  &=\sum\limits_{k=0}^{\infty }{\sum\limits_{i=0}^{\infty }{{{\left( \frac{{{u}_{0}}-1}{{{u}_{0}}} \right)}^{k+i}}\!\!\int_{0}^{\infty }\!\!\!{{{e}^{-ks}}{{l}_{\beta }}\left( s,t \right)\mathrm{d}s\int_{0}^{\infty }\!\!\!{{{e}^{-iz}}{{l}_{\beta }}\left( z,t \right)\mathrm{d}z}}}}\nonumber\\
  &=\int_{0}^{\infty }\!\!\!{\int_{0}^{\infty }{\sum\limits_{k=0}^{\infty }{\sum\limits_{i=0}^{\infty }{{{\left(\frac{{{u}_{0}}-1}{{{u}_{0}}} \right)}^{k+i}}{{e}^{-ks}}{{e}^{-iz}}{{l}_{\beta }}\left( s,t \right){{l}_{\beta }}\left( z,t \right)\mathrm{d}s\mathrm{d}z}}}}\nonumber\\
  &=\int_{0}^{\infty }\!\!\!{\int_{0}^{\infty }\!\!\!{{{u\left( s \right)u\left( z \right){{l}_{\beta }}\left( s,t \right){{l}_{\beta }}\left( z,t \right)\mathrm{d}s\mathrm{d}z}}}}\,.
\end{align}
The substitution of \eqref{eq.SolPowerTwo} for the term ${{w}^{2}}\left( t \right)$ in \eqref{eq.Main} leads to
\begin{align}\label{eq.MainByPart}
  \caputo{w}&=\frac{{{u}_{0}}{{t}^{-\beta }}}{\Gamma (1-\beta )}+w-\int_{0}^{\infty }\!\!{\int_{0}^{\infty }\!\!\!{{{u}^{2}}\left( s \right){{l}_{\beta }}\left( s,t \right){{l}_{\beta }}\left( z,t \right)\mathrm{d}s\mathrm{d}z}}\nonumber\\
  &=\frac{{{u}_{0}}{{t}^{-\beta }}}{\Gamma (1-\beta )}+w-\int_{0}^{\infty }\!\!\!{\left( {{u}^{2}}\left( s \right){{l}_{\beta }}\left( s,t \right)\int_{0}^{\infty }\!\!\!{{{l}_{\beta }}\left( z,t \right)\mathrm{d}z} \right)\mathrm{d}s}\,,
\end{align}
and by referring to \eqref{eq.DelLbeta} and \eqref{eq.ProbInvInf}, it is eventually obtained from the equation \eqref{eq.MainByPart} that
\begin{align}\label{eq.MainByPartCont}
  \caputo{w}&=\frac{{{u}_{0}}{{t}^{-\beta }}}{\Gamma (1-\beta )}+w-\int_{0}^{\infty }\!\!\!{{{u}^{2}}\left( s \right){{l}_{\beta }}\left( s,t \right)\mathrm{d}s}\nonumber\\
  &=\frac{{{u}_{0}}{{t}^{-\beta }}}{\Gamma (1-\beta )}+\int_{0}^{\infty }\!\!\!{\left( u\left( s \right)-{{u}^{2}}\left( s \right) \right){{l}_{\beta }}\left( s,t \right)\mathrm{d}s}\nonumber\\
  &=\frac{{{u}_{0}}{{t}^{-\beta }}}{\Gamma (1-\beta )}+\int_{0}^{\infty }\!\!\!{{u}'\left( s \right){{l}_{\beta }}\left( s,t \right)\mathrm{d}s}\nonumber\\
  &=\frac{{{u}_{0}}{{t}^{-\beta }}}{\Gamma (1-\beta )}+\left. u \left( s \right) {{l}_{\beta }}\left( s,t \right) \right|_{s=0}^{\infty }-\int_{0}^{\infty }\!\!\!{u\left( s \right){{\partial }_{s}}{{l}_{\beta }}\left( s,t \right)\mathrm{d}s}\nonumber\\
  &=\frac{{{u}_{0}}{{t}^{-\beta }}}{\Gamma (1-\beta )}+\left(0-{{u}_{0}}\frac{{{t}^{-\beta }}}{\Gamma \left( 1-\beta  \right)}\right)+\int_{0}^{\infty }\!\!\!{u\left( s \right){}^{C}\!D_{t}^{\beta }{{l}_{\beta }}\left( s,t \right)\mathrm{d}s}\nonumber\\
  &={}^{C}\!D_{t}^{\beta }\int_{0}^{\infty }\!\!\!{u\left( s \right){{l}_{\beta }}\left( s,t \right)\mathrm{d}s}\nonumber\\
  &={}^{C}\!D_{t}^{\beta}w\left(t\right)\,.
\end{align}
\begin{figure}[ht!]
\centering
\includegraphics[scale=.5]{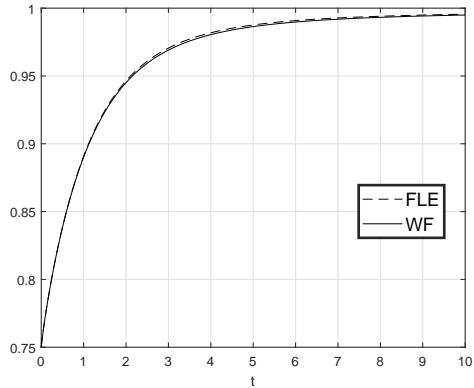}
\caption{Comparison of the West Function (WF) expressed in Eq. \eqref{eq.Solution} and the numerical integration of the FLE (Eq. \eqref{eq.FracLogisticKOne}), for  $\beta = 0.9$ and $u_0=0.75$\,.}
\label{fig2}
\end{figure}
Therefore the function $w\left( t \right)$, expressed in \eqref{eq.Solution}, satisfies the fractional differential equation \eqref{eq.Main}. Figure \ref{fig2} illustrates the graphs of the WF and numerical solution to \eqref{eq.FracLogisticKOne} and shows that the WF is in good agreement with the numerical solution of FLE. Specifically, as mentioned in \cite{west2015}, the WF and numerical solution to FLE coincide for $\beta=1$. As it is obvious from Eq. \eqref{eq.Solution}, the solution to MFLE is obtained by means of a series of Mittag--Leffler functions. Thus, series of Mittag--Leffler functions seem to play an interesting role in the context of fractional logistic equations. The properties of series of Mittag--Leffler functions have been studied in \cite{paneva2010series}.


\section{Estimation of the fractional order}\label{sec.estimation}

The determination of the order of fractional differential equations is an issue, which has been analysed and discussed in recent years \cite{hatano2013,mirko17det} and it has a wide range of applications in physical phenomena such as fractional diffusion equations. In \cite{mirko17det}, fractional order estimation has been conducted for some classes of linear fractional differential equations. In this section, the relationship between the fractional order and the asymptotic behaviour of the solution to MFLE is proved. The solution to \eqref{eq.Main} could be asymptotically expressed by referring to \eqref{eq.AsympNeg}:
\begin{align}\label{eq.SolutionRe}
  w\left( t \right)
  &=\sum\limits_{k=0}^{\infty }{{{\left( \frac{{{u}_{0}}-1}{{{u}_{0}}} \right)}^{k}}{{E}_{\beta }}\left( -k{{t}^{\beta }} \right)}\nonumber\\
  &=1+\sum\limits_{k=1}^{\infty }{{{\left( \frac{{{u}_{0}}-1}{{{u}_{0}}} \right)}^{k}}{{E}_{\beta }}\left( -k{{t}^{\beta }} \right)}\, ,
\end{align}
and for large $t$, by using the Eq. \eqref{eq.AsympNeg}, $w\left(t\right)$ will be approximately equal to
\begin{eqnarray}\label{eq.SolutionAsymOne}
  w\left( t \right) & \approx & 1+\sum\limits_{k\ge 1}{{{\left( \frac{{{u}_{0}}-1}{{{u}_{0}}} \right)}^{k}}\sum\limits_{s\ge 1}{{{\left( -1 \right)}^{s+1}}{{\left( \frac{1}{k{{t}^{\beta }}} \right)}^{s}}\frac{1}{\Gamma \left( 1-\beta s \right)}}}\nonumber\\
  & \approx & 1+\sum\limits_{k\ge 1}{\sum\limits_{s\ge 1}{{{\left( -1 \right)}^{s+1}}{{\left( \frac{{{u}_{0}}-1}{{{u}_{0}}} \right)}^{k}}\frac{{{t}^{-s\beta }}}{{{k}^{s}}}\frac{1}{\Gamma \left( 1-\beta s \right)}}}\nonumber\\
  & \approx & 1+\frac{{{t}^{-\beta }}}{\Gamma \left( 1-\beta  \right)}\sum\limits_{k\ge 1}{{{\left( \frac{{{u}_{0}}-1}{{{u}_{0}}} \right)}^{k}}\frac{1}{k}} \nonumber\\
  & & + \sum\limits_{s\ge 2}{{{\left( -1 \right)}^{s+1}}\frac{{{t}^{-s\beta }}}{\Gamma \left( 1-\beta s \right)}\sum\limits_{k\ge 1}{{{\left( \frac{{{u}_{0}}-1}{{{u}_{0}}} \right)}^{k}}\frac{1}{{{k}^{s}}}}} \, ,
\end{eqnarray}
\begin{remark}
  By observing the Eq. \eqref{eq.SolutionAsymOne}, it is obviously found that the function $w\left( t \right)$ has the limit $ w_{\infty} = 1 $, which is independent of the fractional order, $\beta$, as time tends to infinity.
\end{remark}
\begin{figure}[ht!]
\centering
\includegraphics[scale=.5]{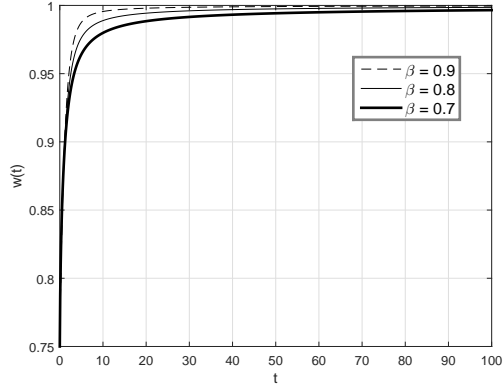}
\caption{The graph of the WF for $\beta = 0.7$, $\beta = 0.8$, $\beta = 0.9$\,.}
\label{fig3}
\end{figure}
Figure \ref{fig3} shows that the solution to MFLE is asymptotically independent of the fractional order, $\beta$, and its limit is equal to one as $t$ goes to infinity. For ${{u}_{0}}\ge \frac{1}{2}$, Eq. \eqref{eq.SolutionAsymOne} is as follows
\begin{eqnarray}\label{eq.SolutionAsymTwo}
  w\left( t \right) & \approx & 1+\frac{{{t}^{-\beta }}}{\Gamma \left( 1-\beta  \right)}\ln {{u}_{0}}+\nonumber\\
  & & \sum\limits_{s\ge 2}{{{\left( -1 \right)}^{s+1}}\frac{{{t}^{-s\beta }}}{\Gamma \left( 1-\beta s \right)}\sum\limits_{k\ge 1}{{{\left( \frac{{{u}_{0}}-1}{{{u}_{0}}} \right)}^{k}}\frac{1}{{{k}^{s}}}}}\, ,\quad{{u}_{0}}\ge \frac{1}{2}
\end{eqnarray}
As $t$ tends to infinity, by neglecting the third term of the right-hand side of \eqref{eq.SolutionAsymTwo}, the function $w\left(t\right)$ is asymptotically equal to
\begin{equation}\label{eq.SolutionAsymFin}
  w(t)\approx 1+  \frac {t^{-\beta} }{\Gamma(1- \beta)}\ln u_0 \, ,\quad{{u}_{0}}\ge \frac{1}{2}\, .
\end{equation}
Therefore, by using the asymptotic behaviour of the function $w\left(t\right)$, the order of the fractional integro--differential equation \eqref{eq.Main} is determined
\begin{equation}\label{eq.OrderEstimate}
  \lim_{t\to +\infty}\, \frac{t{w}'\left( t \right)}{1-w\left( t \right)}=\beta \, .
\end{equation}

\section{Conclusion}\label{sec.Conclusion}

A fractional integro--differential equation is represented, to which the WF expressed in \eqref{eq.Solution} is a solution. The proposed fractional integro--differential equation is called modified fractional logistic equation (MFLE) and its solution is in the form of a series of Mittag--Leffler functions.

\section*{References}

\bibliographystyle{elsarticle-num}
\bibliography{references}

$^1$ Dipartimento di Scienze di Base e Applicate per l'Ingegneria \\
Sapienza Universit\`{a} di Roma \\
Via Antonio Scarpa n. 16 \\
00161 Rome, Italy \\[4pt]
  e-mail: mirko.dovidio@sbai.uniroma1.it
  \\[12pt]

$^2$ Dipartimento di Scienze di Base e Applicate per l'Ingegneria \\
Sapienza Universit\`{a} di Roma \\
Via Antonio Scarpa n. 16 \\
00161 Rome, Italy \\[4pt]
  e-mail: paola.loreti@sbai.uniroma1.it
   \\[12pt]

$^4$ Dipartimento di Scienze di Base e Applicate per l'Ingegneria \\
Sapienza Universit\`{a} di Roma \\
Via Antonio Scarpa n. 16 \\
00161 Rome, Italy \\[4pt]
  e-mail: sima.sarvahrabi@sbai.uniroma1.it

\end{document}